\documentclass[a4paper]{mem}
\usepackage{natbib}
\usepackage{graphicx}
\usepackage[a4paper]{hyperref}
\idline{73}{23}
\begin{document}
\title{Variable stars as tracers of stellar populations in Local Group 
galaxies: Leo~I and NGC~6822}

\author{L. Baldacci \inst{1,2}, 
     F. Matonti \inst{1}, 
     L. Rizzi \inst{3},
     G. Clementini \inst{2},
     E. V. Held \inst{3},
     Y. Momany \inst{4}, 
     L. Di Fabrizio \inst{5},
     and I. Saviane \inst{6} }

\offprints{L. Baldacci}
\mail{via Ranzani 1, 40127 Bologna}

\institute{
 Dipartimento di Astronomia, Universit\`a di Bologna, Bologna, Italy\\
 \and INAF, Osservatorio Astronomico di Bologna, Bologna, Italy\\
 \and INAF, Osservatorio Astronomico di Padova,  Padova, Italy\\
 \and Dipartimento di Astronomia, Universit\`a di Padova, Padova, Italy\\
 \and Telescopio Nazionale Galileo, 38700 Santa Cruz de La Palma,
 Spain \\
 \and European Southern Observatory, Casilla 19001 Santiago, Chile
}

\abstract{Results are presented of a study of the variable star
populations in the dwarf spheroidal galaxy Leo~I and in the dwarf
irregular galaxy NGC~6822, based on time series photometry obtained
with the Wide Field Imager of the 2.2 m ESO/MPI telescope (Leo~I) and
the Very Large Telescope (NGC~6822). We found about 250 (lower limit)
variables in Leo~I most of which are RR Lyrae stars.
In NGC~6822 we identified 450 candidate variables among which
about 20 are RR Lyrae stars, and many are low-luminosity, small-amplitude
Cepheids.
   }
   \authorrunning{L. Baldacci et al.}
   \titlerunning{Variable stars in the Local Group: Leo~I and 
	    NGC~6822}
   \maketitle
%

\section{Introduction}

Variable stars allow to sample different stellar populations in galaxies and
their radial distributions.
The RR Lyrae stars and the Population II Cepheids trace the oldest stellar 
component 
(age $> 10$ Gyr), while the Anomalous and Classical Cepheids trace younger 
population components. 
\section{Leo~I}
The dwarf spheroidal galaxy Leo~I has long been thought to host only
young and intermediate age stellar populations (Lee et al. 1993; Fox
\& Pritchet 1987; Reid \& Mould 1991; Demers, Irwin \& Gambu 1994).
Based on HST WFPC2 data Gallart et al. (1999b) concluded that the bulk
of the stellar formation in Leo~I was delayed, although some hints of
an old stellar population were noted by Gallart et al. (1999a) and
Caputo et al. (1999). 
A delayed first epoch of star formation was in
contrast with the general trend of the Local Group dwarf spheroidal
galaxies, all of which have been found to contain an old stellar
component.
The presence of stars as old as $t > 10$ Gyr remained unproven in
Leo~I until the study by Held et al. (2000), who reported the first
observational evidence of a well extended (from blue to red)
horizontal branch (HB) in the external regions of the galaxy.
A further confirmation was the first detection of a conspicuous
populations of RR Lyrae stars in the galaxy (Held et al. 2001, paper
I).  Paper I focused on the RR Lyrae stars identified in only two
chips of the WFI mosaic. Here we present results on the variables
detected in all the 8 chips of the mosaic.

\begin{figure*}
\centering
\resizebox{12cm}{!}{\includegraphics[bb=144 131 445 711,clip]{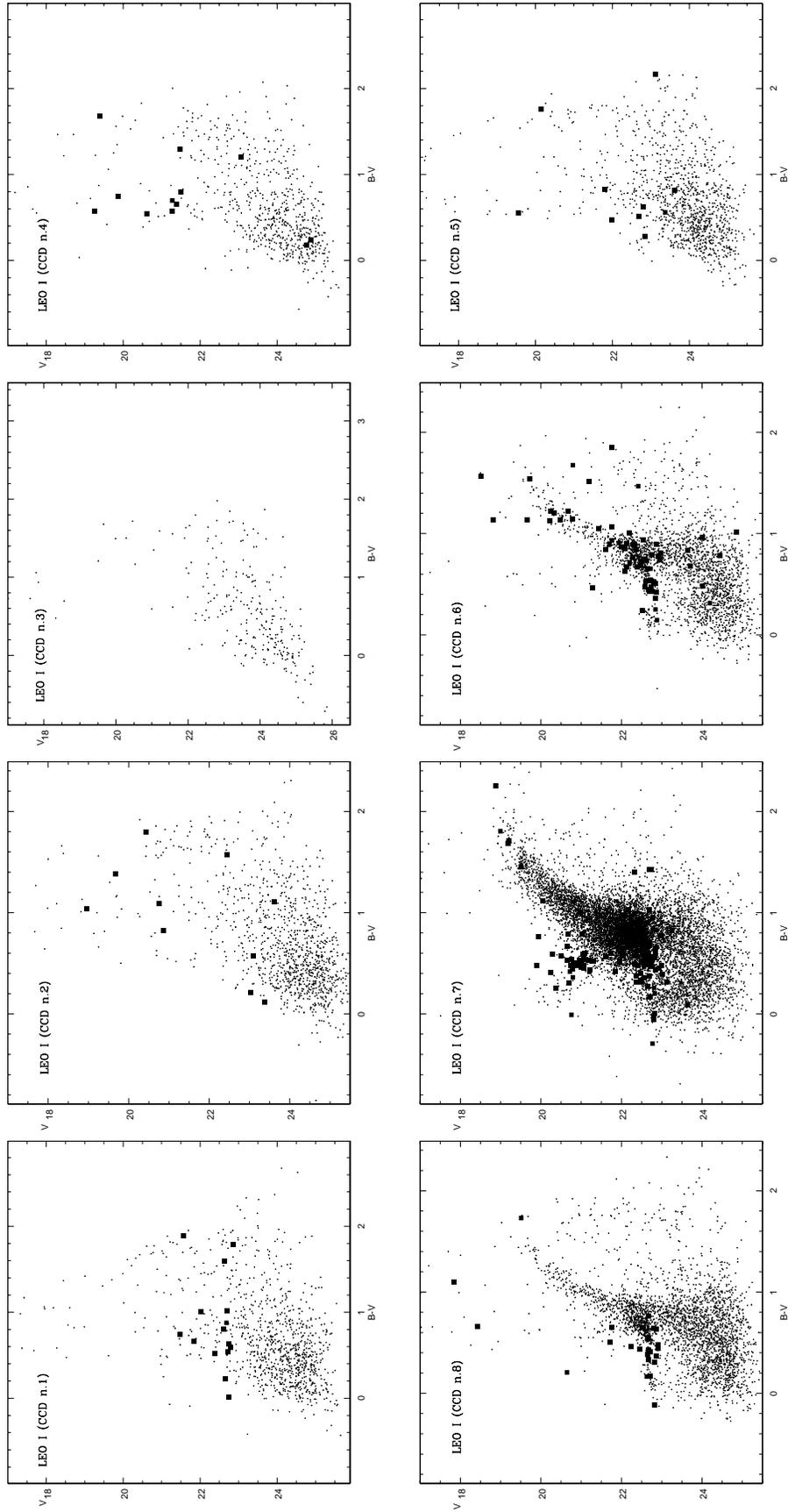}}
\caption{Color-magnitude diagrams of Leo~I, showing the location of 
variable stars (filled squares).}
\label{fig:CMdiagrams}
\end{figure*}

\subsection{Observations and data reduction}
Observations (40 $V$, 22 $B$ and 5 $I$ 15-min exposures) of Leo~I were
obtained with the Wide Field Imager (WFI) at the 2.2m
ESO/MPI telescope. The 34 $\times$ 33 arcmin field of view of the WFI
allows to fully cover the galaxy in just one exposure.
Photometric reductions were performed using DAOPHOT and ALLFRAME
(Stetson 1994), and the WFI reduction pipeline developed by the Padova
group (Rizzi \& Held 2003, in prep.).  
Candidate variables were identified using ISIS2.1 (Alard 2000), a
package based on an optimal image subtraction method.  
In chip~1, where the small number of objects did not allow to run
ISIS2.1, variables were identified using a variability index related to 
the large scatter of their photometric measurements.
Because of telescope pointing drifts during the observations we had to
cut images to a same common area to allow geometrical alignment, 
with a loss of about 1/3 in the sky coverage.  
The total number of candidate variables
identified in Leo~I is $\sim 1000$. However, only about 300 of them
are present in both the $V$ and $B$ master co-added images created by
ISIS.  Thus, a large fraction of the objects flagged as candidate
variables could be spurious identifications.  All candidate variables
are being analyzed with GRATIS, a code developed at the Bologna
Observatory (see Clementini et al. 2000) in order to check their
actual light variation and to determine periods, amplitudes, mean
magnitudes and type classification, whenever possible. The full
catalogue of variable stars in Leo~I will be presented in a
forthcoming paper (Clementini et al. 2003, in prep.).

\subsection{Color Magnitude Diagrams}
Figure \ref{fig:CMdiagrams} shows the color magnitude diagrams of the 
8 individual chips of the mosaic. 
Filled squares mark the position of the candidate variables. 
For a preliminary qualitative analysis, we will consider here only 
chips 6, 7 and 8 where most of the Leo~I stars are located, with
the galaxy center being in chip~7. 
The color magnitude diagram of chip~7 is mainly populated by an intermediate  
age population. The red giant branch (RGB) and the red clump stars 
are the most visible structures. 
The Leo~I HB is completely hidden by the main sequence, 
subgiant branch and red clump star populations, and can be recognized  
thanks to the RR Lyrae stars (see Held et al. 2001).
Several other candidate variables 
are present in the CMD. 
The most 
striking feature is a 
group of variables found in the strip 
about 1.4 mag brighter than the HB.
Their position is consistent with the location in the HR diagram of the
intermediate mass helim-burning pulsating stars that have been found in several
Local Group dwarf spheroidal galaxies (see Pritzl et al. 2002, and references
therein) and that are generally referred to as Anomalous Cepheids (Wallerstein
\& Cox, 1984).
On the other hand, it seems unlikely that these stars could be Population II 
Cepheids, i.e. post HB stars (t$>$ 10 Gyr, $\rm{M\leq 1M_{\odot}}$)
coming from the bluer part 
of the HB and crossing the instability strip in their evolution 
to the Asymptotic Giant Branch. Population II Cepheids are commonly found
in globular clusters with 
extended blue tails, but    
have rarely been found in dwarf galaxies. Moreover, Held et al. (2000) 
showed that the blue HB of Leo~I 
is less extended than that 
of M5.
%
Chip~6 and 8 can roughly be considered as representative of the outer regions 
of Leo~I.
Their CMDs look very similar 
to each other, and are 
different from that of chip~7. The RGBs are 
narrower than in chip~7 and the HBs are clearly visible and populated
by several RR Lyrae stars. 
Conversely,  
the two diagrams do not show evidence for a conspicuous population
of variables above the HB.
All these considerations suggest that in Leo~I the old population is
present all the way through the inner regions of the galaxy, while the 
younger populations, traced by the variables
1.4 mag brighter than RR 
Lyrae stars, are more centrally concentrated. 
Such a trend is commonly observed 
in many dwarf galaxies (Held et al. 2000, and references 
therein). 

\section{NGC~6822}
The star formation history of the dwarf irregular galaxy NGC~6822 has been
investigated by Gallart (1996a,b) and Wyder (2001, 2003) and found to  be 
consistent with both a first epoch of star formation about 10 Gyr ago  
and a star formation started about 6 Gyr ago. 
Here we briefly summarize our results on the search for short period
variable stars in NGC~6822, a more extended discussion being presented
in Clementini et al. (2003) and in a forthcoming paper 
(Held et al. 2003, in prep.). 
Based on VLT time series observations (36 $V$ and 11 $B$ frames) we
detected 450 candidate variables in a region covering about 1/4 of the
galaxy. Among the subsample of faint variables ($V > 23$ mag) we
found a few ($\sim$ 20) RR Lyrae stars tracing the galaxy's HB, and
several Cepheids characterized by low luminosities (LL Cepheids, a few
tenth of magnitude brighter than the HB) and small amplitudes (0.1 -
0.4 mag). The detection of RR Lyrae stars in NGC~6822 breaks the
degeneracy in the galaxy star formation history scenarios, providing
indisputable evidence that NGC~6822 started forming stars at least 10
Gyr ago, and allowing us to determine the galaxy distance using for the
first time a Pop. II distance indicator: $(m - M)_0$=23.36 $\pm$0.17
mag (Clementini et al. 2003).  The nature of the many small-amplitude
LL Cepheids detected in NGC~6822 is still unclear, but theoretical
models of Anomalous Cepheids predict objects with luminosities and
amplitudes similar to those we have detected in NGC~6822 (Fiorentino
et al.  2003, Caputo et al. 2003 in prep.).


\bibliographystyle{aa}

\end{document}